# Photonics for artificial intelligence and neuromorphic computing

Bhavin J. Shastri[a,b,g,h], Alexander N. Tait[c,b,g,h], Thomas Ferreira de Lima[b], Wolfram H. P. Pernice[d], Harish Bhaskaran[e], C. David Wright[f], Paul R. Prucnal[b]

[a]Department of Physics, Engineering Physics & Astronomy, Queen's University, Kingston, ON KL7 3N6, Canada
[b]Department of Electrical Engineering, Princeton University, Princeton, NJ 08544, USA
[c]Applied Physics Division, National Institute of Standards and Technology, Boulder, CO 80305, USA
[d]Institute of Physics, University of Muenster, Muenster 48149, Germany
[e]Department of Materials, University of Oxford, Oxford OX1 3PH, UK
[f]Department of Engineering, University of Exeter, Exeter EX4 4QF, UK
[g]These authors contributed equally to this work.
[h]shastri@ieee.org; alexander.tait@nist.gov

**Research in photonic computing has flourished due to the proliferation of optoelectronic components on photonic integration platforms. Photonic integrated circuits have enabled ultrafast artificial neural networks, providing a framework for a new class of information processing machines. Algorithms running on such hardware have the potential to address the growing demand for machine learning and artificial intelligence, in areas such as medical diagnosis, telecommunications, and high-performance and scientific computing. In parallel, the development of neuromorphic electronics has highlighted challenges in that domain, in particular, related to processor latency. Neuromorphic photonics offers sub-nanosecond latencies, providing a complementary opportunity to extend the domain of artificial intelligence. Here, we review recent advances in integrated photonic neuromorphic systems, discuss current and future challenges, and outline the advances in science and technology needed to meet those challenges.**

Conventional computers are organized around a centralized processing architecture (i.e. with a central processor and memory), which is suited to run sequential, digital, procedure-based programs. Such an architecture is inefficient for computational models that are distributed, massively parallel, and adaptive, most notably, those used for neural networks in artificial intelligence (AI). AI is an attempt to approach human level accuracy on these tasks that are challenging for traditional computers but easy for humans. Major achievements have been realized by machine learning (ML) algorithms based on neural networks [1], which process information in a distributed fashion and adapt to past inputs rather than being explicitly designed by a programmer. ML has had an impact on many aspects of our lives with applications ranging from translating languages [2] to cancer diagnosis [3]. Neuromorphic engineering is partly an attempt to move elements of ML and AI algorithms to hardware that reflects their massively distributed nature. Matching hardware to algorithms leads potentially to faster and more energy efficient information processing. Neuromorphic hardware is also applied to problems outside of ML, such as robot control, mathematical programming, and neuroscientific hypothesis testing [4, 5]. Massively distributed hardware relies heavily—more so than other computer architectures—on massively parallel interconnections between lumped elements (i.e. neurons). Dedicated metal wiring for every connection is not practical. Therefore, current state-of-the-art neuromorphic electronics use some form of shared digital communication bus that is time-division multiplexed, trading bandwidth for interconnectivity [4]. Optical interconnects could negate this trade-off and thus have the potential to accelerate ML and neuromorphic computing.

Light is established as the communication medium of telecom and datacenters, but it has not yet found widespread use in information processing and computing. The same properties that allow optoelectronic components to excel at communication are at odds with the requirements of digital gates [6]. However, non-digital computing models, such as neural networks, can be more conducive to being implemented in photonics. The goal of neuromorphic photonic processors should not be to replace conventional computers but to enable applications that are currently unreachable by conventional computing technology, specifically, those requiring low latency, high bandwidth, and low energies [7]. Examples of applications for ultrafast neural networks include:

- Enabling fundamental physics breakthroughs: qubit readout classification [8], high-energy particle collision classification [9, 10], fusion reactor plasma control [11]
- Nonlinear programming: solving nonlinear optimization problems (robotics, autonomous vehicles, predictive control) [12], and partial differential equations [13]

- Machine learning acceleration: vector-matrix multiplications [14], deep learning inference [15], ultrafast or online learning [16]
- Intelligent signal processing: wideband RF signal processing [17], fiber-optic communication [18, 19]

Photonic circuits are well suited to high-performance implementations of neural networks for two predominant reasons: interconnectivity and linear operations. Connections between pairs of artificial neurons are described by a scalar synaptic weight (a primary memory element), so the layout of interconnections can be represented as a matrix-vector operation, where the input to each neuron is the dot product of the output from connected neurons attenuated by a weight vector. Optical signals can be multiplied by transmission through tunable waveguide elements, and they can be added through wavelength-division multiplexing (WDM), by accumulation of carriers in semiconductors [20,21], electronic currents [22, 23], or changes in the crystal structure of material induced by photons [24]. Neural networks require relatively long-range connections to perform non-trivial distributed information processing. When comparing metal wire connections to photonic waveguides, as a function of distance, optical signals experience lower attenuation and generate less heat (the latter provided the light source if off-chip). More importantly, waveguides have no inductance or skin effect, which means that frequency-dependent signal distortions are minimal for the long-range connections present in neural interconnects. Electronic point-to-point links today take advantage of transmission line and active buffering techniques; however, neural networks are not based on point-to-point links but instead involve massively parallel signal fan-out and fan-in. It is not practical to use state-of-the-art transmission line and active buffering techniques for each physical connection. Consequently, to avoid the tradeoffs exhibited by electronic wiring, neuromorphic electronic architectures employ digital time-multiplexing [4] that allows for the construction of larger neural networks at the expense of bandwidth. For many applications though, bandwidth and latency are paramount, and these applications can be met only by direct, non-digital photonic broadcast (i.e. many-to-many) interconnects.

Optics has long been recognized as a promising medium for matrix multiplication [25] and interconnects [26, 27]. Optical approaches to neural networks were pioneered decades ago by Psaltis and others [28]. Today, societal demands for computing have changed. It is this new demand as well as factors related to maturity of enabling technologies that has created a renewed case for photonic neural networks. One factor is silicon photonics, which is a crucial advance over previous efforts. Silicon photonic platforms can host high-quality passive components combined with high-speed active optoelectronics, all available with competitive integration density [29]. In 2014, some of us first introduced a proposal for a scalable silicon photonic neural network [21], which was demonstrated in 2017 [30] concurrently with other silicon photonic neuromorphic architectures [14, 22]. On-chip silicon electronics for calibration and control provide a route to overcome component sensitivity [31], and progress in on-chip optoelectronics provides a route to cascadability and nonlinearity [32]. The ability of neuromorphic photonic systems to provide step-changes in our computing capabilities is moving ever closer, with, potentially, PetaMAC(multiply-accumulate operations) per second per $mm^2$ processing speeds [33] and attojoule per MAC energy efficiencies [32]. While photonics provides advantages in connectivity and linear operations over electronics, other aspects, such as storing and accessing neuron weights in on-chip memory, present new challenges. There has been significant investigations on optical memories including 'in-memory' computing [34, 35], however they cannot usually be written to and read from at high frequencies. Future scalable neuromorphic photonic processors will need to have a tight co-integration of electronics with potentially hybrid electronic and optical memory architectures, and take advantage of the memory type (volatile vs non-volatile) in either digital or analogue domains depending on the application and the computation been performed.

In this review, we survey recent research and current challenges in neuromorphic photonics and highlight possible solutions. We discuss analog interconnects—an area where optics conventionally excels—and approaches for implementing neuron nonlinearities—an area where optics conventionally faces difficulties. Neural networks fall into a number of general categories: layered or recurrent, spiking or continuous-time, etc. We survey these various categories and their implications for hardware implementation. The survey directs readers to prior reviews of photonic reservoir computing, a related area of photonic information processing. In the subsequent sections, we discuss key technologies needed for demonstrations of neuromorphic photonic hardware to scale to practical systems, including active on-chip electronics, light sources. Finally, we highlight some emerging research directions towards increasing functionality and efficiency, including non-volatile memory, photonic digital-to-analog converters (DACs), and frequency comb sources.

## Survey of photonic neural networks

Research in neuromorphic photonics encompasses a variety of hardware implementations, and, crucially, multiple neural network types, each with different application classes. In general, all types of neural networks consist of nonlinear elements (a.k.a. neurons) interconnected via configurable, linear weights (a.k.a. synapses)–see Box 1. Each neural model has a different signal representation, training method, and network topology. For example, artificial neurons with a continuous-variable nonlinear transfer function can be trained with backpropagation via gradient descent [1, 36], while spiking neurons are better suited to different spike-time-dependent update rules.

**Implementations of weighted interconnects (synapses).** Connections between a pair of neurons are weighted by their intervening synapse. These synaptic weights are scalar multipliers. Before being received by downstream neurons, the weighted signals from upstream neurons are summed. The weighted interconnects can therefore be represented by a matrix whose entries are the weight values, with each entry multiplying a particular synapse's input signal. One purpose of the photonic system is to perform that matrix multiplication.

Figure 1 shows demonstrations of various integrated photonic circuits for matrix multiplication and weighted interconnection. Implementations fall into two broad categories, one based on wavelength and the other on optical modes. In Fig. 1a, WDM signals are weighted in parallel by a bank of microring resonators (MRRs) used as tunable filters [37, 38]. This approach forms an essential part of an integrated architecture called "broadcast-and-weight," first proposed in [21] and demonstrated in [30] (see also Fig. 4a). Several other architectures for multiwavelength synapses and neural networks have been proposed. Most employ WDM fan-in for weighted addition [23, 24, 39, 40], but they differ in terms of how the channels are weighted. The architectures in refs. [24, 39] demultiplex the wavelengths, attenuate each channel, and then remultiplex before WDM fan-in. In Ref. [39] (Fig. 1b), the weighted attenuators are made from semiconductor optical amplifiers (SOAs); in Ref. [24] (see also Fig. 2g), the tunable attenuators are composed of phase change materials (PCMs).

An array of beam splitters and phase shifters can implement unitary matrix transforms using interference between different paths of coherent input light [41], where inputs are assigned to different waveguides and power modulated. This principle underlies mode-based weighted interconnection for photonic neural networks, of which examples are shown in Fig. 1c. This unitary transformation architecture was implemented on an integrated platform using thermally tuned silicon waveguides and directional couplers arranged in a mesh of Mach-Zehnder interferometers (MZIs) [42]. Neural network interconnects can be any matrix—not just unitary. The needed non-unitary neural interconnect was shown to be possible by factoring the weight matrix into one unitary MZI-mesh, one array of tunable attenuators, and a second unitary MZI-mesh [14] (Fig. 1c). The cryogenic architecture from ref. [22, 43] uses different optical modes in multiple waveguide layers [44] (Fig. 1d). In contrast to the MZI-mesh, signals are incoherent and produced by all-silicon integrated light sources [45]. In addition to the dense crossbar layout, the multi-layered waveguide approach enables complex waveguide routing layouts [43].

Modulation of the effective refractive index of signal-carrying waveguides is another optical mode-based approach to weight configuration. A number of index tuning mechanisms have been developed. Thermal tuning with metal filament microheaters, is perhaps the easiest way to effect large index changes, but it is slow and power inefficient [46]. Thermal tuning with waveguide-embedded heaters is similar [47], but provides a feedback signal for weight control [48]. In silicon, the strongest effects are the thermooptic effect, free-carrier absorption and free-carrier dispersion (also known as plasma dispersion). One can directly manipulate carrier concentrations by selectively p- and n-doping the waveguide in a lateral junction [49]. Alternatively, hybrid waveguides can be made of a silicon core and other materials with favorable index modulation properties close enough to the core that interact with the evanescent field. Some examples include III-V hybrid integration [50], lithium niobate [51], and graphene [52] modulation. These mechanisms are faster and require much less power compared to heaters, but typically provide smaller tuning range before electrical damage. Tuning methods based on chalcogenide PCMs allow weights to retain their values without further holding power after being set [24, 35].

Examples of non-volatile synapse implementations are shown in Fig. 1e and 1f. These materials have been referred to as "all-optical" because they do not need electrical inputs for tuning. Both are based on the use of optically induced changes in chalcogenide materials to control the light propagation in waveguides (the former $Si_3N_4$ integrated waveguides [24], the latter metal-sulphide fibers [53]). Weight configuration based on non-volatile optical materials could have significant impact on the challenges of electrical I/O and heat dissipation.

**Implementations of nonlinearities (neurons).** In all neural network models, some form of nonlinearity is required in the primary signal pathway in order to implement the thresholding effect of the neuron. A multitude of photonic devices exhibit nonlinear transfer functions that resemble neuron-like or gate-like transfer functions; however, a nonlinear response alone is not sufficient for a photonic device to act as a neuron. Photonic neurons must be capable of reacting to multiple optical inputs (a.k.a. fan-in), applying a nonlinearity, and producing an optical output suitable to drive other like photonic neurons (a.k.a. cascadability). Optical devices face fundamental challenges in satisfying these requirements in particular, as pointed out by Keyes and Goodman decades ago [6,54]. Today, these challenges and requirements are being addressed with integrated photonic solutions, with some successful approaches shown in Fig. 2. These approaches fall into two major categories based on the physical representation of signals within the neuron: optical-electrical-optical (O/E/O) vs. all-optical.

O/E/O neurons, proposed in 2013 by Nahmias et al. [55] and Romeira et al. [56], involve a transduction of optical power into electrical current and back within the primary signal pathway. The primary signal pathway or drive chain refers to elements representing the rapidly changing neuron state variables akin to membrane voltage or synaptic state. Outside of this primary pathway, one always finds electronics representing the relatively slowly varying neuron parameters and control logic. In O/E/O neuron signal pathways, nonlinearities occur in the electronic domain or in the E/O conversion stage using lasers or saturated modulators. The use of E/O nonlinearities for photonic neurons was shown, using modulators, in [57–59] (Fig. 2a,b) and, using lasers, in [60] (Fig. 2c). A photodetector-modulator neuron for MZI meshes was proposed in [61] (Fig. 2d). Other O/E/O approaches implement nonlinearity purely in the electronic domain, for example, refs. [22, 62] proposed that the nonlinear dynamics of spiking photonic neurons could be implemented with a superconducting electronic signal pathway (Fig. 2e).

In separating input light-matter interaction (O/E) from output light-matter interaction (E/O), O/E/O neurons can modulate their output signal onto a fresh optical carrier that is unconstrained by the optical characteristics (i.e. power, phase, mode, wavelength) of its inputs. Importantly, this means that the output can be significantly stronger than the input, which is not generally the case for all-optical neurons, discussed below. A key distinction among O/E/O photonic neurons is whether the high-bandwidth nonlinear transfer function is imparted by analog components (e.g. E/O responses, transistor circuits, single-photon detectors) vs. by a digital lookup circuit [14, 39, 61]. The transfer functions of analog neurons can be configured by electrical biasing, but their shapes are constrained by the response of whatever device provides the nonlinearity. Digital counterparts provide the flexibility to implement arbitrary transfer functions. Several works have shown, however, that matching particular transfer function shapes is not necessary: neural training and programming techniques can be adapted to transfer functions naturally exhibited by analog photonic devices [18, 30, 63].

All-optical neurons do not ever represent the neuron signal as an electrical current but, instead, as changes in material properties such as semiconductor carriers or optical susceptibility. Optical nonlinear susceptibilities are power inefficient—not just very weak—meaning that neuron output is necessarily, often significantly, weaker than its input and thus incapable of driving even a single other neuron. Solutions to fan-out and cascadability have been demonstrated by combining nonlinear optical devices with optical carrier regeneration. Regeneration means that each neuron outputs a fresh carrier wave, which is power modulated by that neuron's output signal, a function of the sum of its input signals. Carrier regeneration approaches were first adapted to photonic neurons in 2002 by Hill et al. [64] and have been used in all experimental demonstrations of photonic neurons to date, including all-optical neurons.

Carrier regeneration involves the control of output light with input light. All-optical neurons must provide this function in addition to a nonlinear function. Carrier regeneration enabling photonic neurons has been shown in a feedforward fashion using semiconductor carrier populations: using cross-gain modulation [65] (Fig. 2f) or using cross-phase modulation in an interferometer [20]. It can also be achieved by changing a material state, such as via a structural phase transition [24, 66] (Fig. 2g). All-optical carrier regeneration introduces a new challenge: differentiating the controller signal from the controlled signal. Both affect the material substrate, so the output must be weaker than the input. Optical amplifiers can be employed to boost the output such that it can drive downstream neurons. Like the continuous-time neurons above, spiking laser neurons can also be categorized into the two broad classes of O/E/O and all-optical. A range of implementations for both classes is summarized in Fig. 3. Spiking laser neurons achieve strong nonlinearity, carrier regeneration, and neural dynamics all within a single device consisting of gain, a cavity, and a saturable process. Spiking neurons have been demonstrated using saturable semiconductor media [55, 67, 68] (Fig. 3a,f), resonant tunneling diodes [56, 69] (Fig. 3b), graphene saturable absorbers [70] (Fig. 3d), and mode competition [71–73] (Fig. 3c,e).

A perceived advantage of all-optical neuron implementations is that they are inherently faster than O/E/O implementations due to relatively slow carrier drift and/or current flow stages in the latter. Indeed, for optical telecommunications, O/E/O for the purpose of digital regeneration is considered undesirable and inefficient; however, the bottleneck is not due to the transduction between light and current—it is caused by a need to demultiplex and digitize many different channels. The majority of proposed O/E/O neurons do not involve digitization, and, thus, rarely impose bandwidth bottlenecks. In fact, recent analog O/E/O devices have exhibited bandwidth and energy performance on par with or better than all-optical components, as compellingly illustrated in the O/E/O work of Nozaki and others [32]. For neurons that apply nonlinearities in the digital domain—the most extreme example being in a CPU—the digital subsystem is often the determinant of maximum system bandwidth.

**Neuromorphic architectures (neural networks).** Neuromorphic hardware architecture is governed by models of artificial neural networks; however, neural networks models have many subclasses. They can differ in terms of neuron signal representation, weight configuration, and network topology. Weight configuration, in this context, refers to the approach of setting weights so that the network accomplishes a particular computational task. Configuration can be guided by supervised training, unsupervised learning (a.k.a. plasticity), or programmatic "compilation." Topology describes the graph structure of non-zero weights between neurons. In the most general case—the all-to-all recurrent topology—there are forward and backward directed connections between each pair of neurons. However, constraining the topology in particular ways can unlock powerful analytical tools to guide weight configuration. For example, feedforward topologies yield to chain rule decomposition [36], and symmetric topologies yield to an energy surface formulation [74]. Configuration and topology are intertwined with signal representation and the behavior of individual neurons. In feedforward networks, the output is completely determined by the present inputs, meaning neuron input signals can be scalar values as opposed to functions of time (although they change when new inputs are presented or when weights are updated). In recurrent networks, on the other hand, the outputs depend also on the history of inputs. This means that neurons must have internal states that evolve in time non-instantaneously. They are referred to as stateful neurons. Box 1 shows two types of stateful neurons, one with continuous-valued outputs, and another with outputs consisting of temporal delta functions (i.e., spikes).

Examples of photonic architectures experimentally demonstrated to date are shown in Fig. 4. The model in Fig. 4a [30] is recurrent, continuous-time, and programmed by compiler [75]. The model used in Fig. 4b [14] is feedforward, single valued, and externally trained. The model in Fig. 4c [24] is feedforward, spiking, with both external and local training. Figure 4d shows a feedforward multilayer perceptron architecture that combines semiconducting few-photon LEDs with superconducting-nanowire single-photon detectors to behave as spiking neurons, connected by a network of optical waveguides [22]. Figures 4e and 4f show free-space diffractive network implementations, the former a recurrent type [76] and the latter a feedforward deep neural network implementation [77]. A free-space diffractive network with a nonlinear activation function has also been demonstrated [78]. The computational tasks of the above examples range over, respectively, audio classification, dynamical system emulation, image classification, nonlinear optimization, and neuroscientific hypothesis testing.

Classification of unstructured signals is a key application of machine learning. Networks used for classification are often feedforward: arranged in layers with all connections from a given layer projecting to the following layer. They are often called "deep networks." Convolutional neural networks (CNNs or ConvNet) [36] are a type of deep networks that consists of a series of convolutional layers that perform a sliding dot product on vectors from the previous layer for local feature extraction, followed by a pooling layer to merge similar features into one. In supervised training, the outputs are compared to the correct outputs for corresponding inputs. The classification error determines how the weights are updated. After this training phase, the network is able to generalize to inputs that have not been seen before. An important case of supervised training for deep networks, called backpropagation, provides rules to update all of the weights in the network [1]. Backpropagation training calculations can be performed by a computer and then applied back to weights and biases in photonic hardware, either coherent approaches in feedforward networks [14, 77, 79] or WDM-based CNNs [23, 80]. Convolutional operations require a delicate balance between processing and memory access—a challenge for photonics discussed later. Photonic hardware implementations of backpropagation have also been proposed [16].

In addition to machine learning and classification, neural networks can be programmed to solve challenging mathematical problems, specifically nonlinear differential equations and nonlinear optimization. Such methods have been applied to photonics to address predictive control for fast moving bodies [12] and Ising machines for the study of many-body physics and molecular chemistry [81, 82]. Networks with inhibitory connections (weights) have also been shown to solve combinatorial problems for compressed sensing [83]. The Neural Engineering Framework (NEF) [75] is a promising tool that mixes compiling and reward-based adaptation to implement block-diagram-style control and decision systems. The NEF represents variables with populations of neurons—as opposed to single neurons—and has been applied to photonic neural networks in [30]. Further work is needed to identify benchmarks and applications of extreme computing with ultrafast neuromorphic hardware [84].

Other architectures under investigation typically involve a higher level of brain-inspired principles. Spiking neuron models represent signals as trains of Dirac delta functions, digital in amplitude but analog in time. Temporal coding (i.e. neural processing related to the timing of spikes) has been proposed as one of the keys to energy efficiency in biological nervous systems [85]. Significant research in photonics has been directed to optoelectronic devices that approximate spiking dynamics [22, 24, 55, 56, 67, 70–72, 86–88]. Spike coding opens key new unsupervised plasticity methods, especially spike-time dependent plasticity (STDP). STDP is considered essential to a number of brain areas and has been explored in photonics in [24, 43, 89–91].

**Photonic reservoir computing.** Photonic reservoir computing (RC) is a growing field of interest and importance in photonic information processing. A full coverage of this area falls outside the scope of this review. The interested reader is directed towards the many excellent reports in the literature (e.g. [92–96] including recent extensive reviews [97, 98]). Reservoir computers consist of a network of random connections and nonlinearities (the reservoir) followed by a readout layer. The reservoir generates a large number of complex possible behaviors in response to an input, and the readout layer is trained to select the behavior that solves a particular computational problem of interest. Many types of physical substrates can be used as the reservoir, including a variety of photonic systems. Research in photonic RC has realized dynamical complexity by combining optical delays with one of either a photodetector and modulator [92], optical amplifiers [94], a semiconductor nonlinearity [95], or a recurrent interferometer mesh [96]. While both reservoir and neuromorphic approaches share roots in neuroscience, they represent complementary approaches to information processing. All neuromorphic systems require a known isomorphism between the physical hardware and a neural network model, upon which training and programming are based; reservoir systems can accomplish processing goals without requiring this isomorphism, instead requiring supervised training to read out desired behaviors. Consequently, despite their behavioral complexity, photonic reservoirs can be simple to construct and have shown rapid progress in task-based demonstrations, for example, a single optoelectronic modulator and optical fiber to classify spoken digits [95], and a spatial light modulator based RC [76] for computer vision tasks to recognize human actions [99].

**Towards a neuromorphic photonic processor**

Mainstream silicon photonic platforms offer a device library (modulators, waveguides, detectors) to implement the main signal pathways in at least some neuromorphic architectures. Modifications to standard manufacturing processes, including the introduction of PCMs and/or superconducting electronics, extend potential realizable architectures even further, as will be discussed in the emerging ideas section. In all architectures, there is a need for complex on-chip electronic circuitry for calibration and control of the network parameters, and there is a need to generate light. Such components are not yet widely available on current commercial silicon photonic platforms. Here, we discuss different routes to integrating the electronic circuits and light sources necessary for a neuromorphic photonic processor. A complete processor in a package is rendered in Fig. 5, illustrating the roles of the respective technologies.

**Active on-chip electronics**. Photonic chips require DC analog signals (e.g. bias voltages/currents), control systems (e.g. feedback, algorithms etc.), interfaces with electronics (e.g. digital-to-analog converters (DACs) and analog-to-digital converter (ADCs)), and require stabilization (e.g. temperature). Dedicated analog electronic circuits are needed for these purposes—some low-bandwidth (DC configuration) and some high-bandwidth (DACs, ADCs, transimpedance amplifiers, feedback). As a result, neuromorphic photonic chips generally require significantly more electrical ports than optical ones, leading to a high electronic interconnect density. Indeed, the required number of electrical ports usually scales quadratically with the number of optical ports. This challenge can be overcome by co-integration of CMOS with photonic chips. There are several technological routes with different tradeoffs. Here, we discuss three of these options: wirebonding, flip-chip bonding, and monolithic fabrication.

CMOS chips with digitally-controlled analog devices can be connected with the photonic integrated circuit (PIC) via lateral wirebonds. However, as the processors scale in number of elements it will (i) become physically impossible to have dedicated wire bonds, with an eventual limit to the electrical I/O, and (ii) be very expensive to have a large part of the chip's real estate dedicated to routing metal traces to its perimeter. In addition, wirebonding imposes a limit on signal bandwidth because of parasitic inductance in the wires. Wirebonding remains, however, a good approach to engineer small-scale prototype systems in a laboratory setting.

Flip-chip bonding involves fabricating two dies, one optimized for CMOS electronics and the other optimized for silicon photonics. The dies have matching electrical pads. They are soldered, or bonded, to one another to create a large number of electrical connections [31]. Fig. 5 illustrates the flip-chip bonding approach. A custom CMOS ASIC mates with the pads on a silicon PIC and provides the interface between the many distributed, continuously configurable optical devices and the predominantly serial, digital realm. Each DAC for each tuning element must be capable of holding its value in a digital register – in other words, the memory for parameters (i.e. weights, biases) is distributed and co-located with corresponding optical elements.

Flip-chip bonding brings several advantages over wirebonding. First, the connection number grows with the area of each chip. Second, the interconnections have reduced parasitic impedance and higher bandwidth. Finally, the multi-die approach can combine the best of electronic and photonic technologies, because each die can be optimized independently. From an economic standpoint, this type of integration does not require new process development and can rely on existing designs for microelectronic microprocessors, memory, and digital-to-analog converters. Multi-die approaches introduce new challenges in managing thermal fluctuations from the CMOS die. Through-silicon via (TSV), a vertical electrical connection that passes through the silicon wafer or die is an alternative strategy to connect electronic and silicon dies. TSVs trade off high integration density and low interconnect parasitics with complexity in thermal management. A comprehensive review and comparison of these technologies is provided in [31].

Monolithic fabrication entails the integration of electronics and photonics on the same substrate, and several approaches have been investigated to achieve this. So-called 'zero-change' platforms earn their name by offering a monolithic fabrication process that is minimally changed from an industry-standard fabrication process (such as a 45 nm SOI (silicon on insulator) CMOS process [100, 101]). Another direction is to adapt SOI photonic processes to enable some active integrated electronic logic [102]. A dedicated homogeneous process called *9 WG*, recently offered by GlobalFoundries [103], looks to balance photonic performance with electronic circuitry for analog or digital control.

**Light sources.** A substantial challenge for silicon as a photonic platform is its inability to generate light on-chip. Current approaches in the silicon photonics data communications industry rely on fiber packaging with external light sources. For neuromorphic silicon photonic systems, fiber packaging is also the most straightforward way to get light into the chip. However, since photonic processing systems do not intrinsically need to send optical signals off-chip, ideally, all optical signals would be confined within an integrated circuit package. Thus, co-packaged light sources will eventually be critical for the efficiency, stability, and scalability of neuromorphic photonics. Fig. 5 depicts both external fiber optic light source and a multi-die approach using a non-silicon die for light generation. Both are shown with photonic wirebond connections, although other techniques for coupling light between dies are available. Substantial research has been dedicated to integrating light sources directly onto the silicon waveguide layer [104]. Approaches include rare-earth element doping, strain engineering of germanium, and all-silicon emissive defects [45]. Each approach offers distinct features, and drawbacks, to thermal efficiency, integration compatibility, scalability, and temperature stability.

Another approach involves the use of III-V devices integrated with silicon. InGaAsP, for example, provides high gain and saturation powers, but is difficult to integrate directly on silicon due to the crystal lattice mismatch between the materials. Ways around this include either (i) using a separate III-V die to couple light to the silicon die, or (ii) integrating III-V's on silicon waveguides. The former has been achieved by pick-and-place bonding of III-V and Si dies with finished gain ridges [105], or using laser micropackages [106]. Pick-and-place techniques can suffer from tight alignment tolerances. System-in-package approaches—where the laser die is placed next to the silicon die—can be enabled by photonic wirebonding [107]. Photonic wirebonding involves writing 3D waveguides in a photosensitive polymer. Since this step is performed after die placement, alignment requirements are relaxed. Integrating III-V's with Si has been achieved by bonding wafers with III-V quantum wells (QWs) to silicon waveguides [108]. Finally, III-V quantum dots (QDs) have been grown directly on silicon [109]. Unlike QW layers, QDs can tolerate lattice mismatch without the loss of their optical gain properties.

The requirements for lasers for neuromorphic photonics depend mostly on the type of neuron. Multi-die techniques are well-suited to neuromorphic systems with modulator-class neurons because the light source can be well outside of the neural signal pathway. For laser-class neurons, gain must be tightly integrated on waveguides, necessitating wafer bonding, QD growth, or silicon emissive defects. Sources for modulator neurons are faced with one of two nontrivial performance requirements. For multiwavelength architectures, as in Fig 1a, many different wavelengths would require many different sources. Another option would be a single source producing multiple wavelengths (e.g. a frequency comb source). Coherent architectures, as in Fig 1c, face almost the opposite challenge: there can only be a single source laser because the network requires a single optical phase reference. A challenge is that this single laser must generate enough optical power for the entire system.

## Emerging ideas and outlook

As pointed out in a recently published 'roadmap' on emerging hardware and technology for machine learning [110], neuromorphic photonics could provide an outstanding candidate. Some neuromorphic photonic architectures are feasible to build on commercial silicon photonic platforms, given electronic and light source integration; however, in order to transform early system demonstrations into practical and fully performant AI processors, neuromorphic photonic systems must evolve to incorporate novel technologies. For example, correcting fabrication variability can reduce heat dissipation and the amount of current needed for tuning. Memory circuits that are able to interact directly with light can enable more agile reconfiguration in the processor. Finally, information can be transferred between electrons and photons with significantly reduced heat dissipation, which can be achieved with further technological improvements in optical sources, high-efficiency modulators, and photonic analog-to-digital interfaces. This section highlights key emerging technologies that could have large impact on the performance potential of neuromorphic photonic processors.

**Memory in neuromorphic processors.** So far, the approaches described here rely on a combination of specialized photonic devices being driven by more generalized electronic circuits or micro-controllers. This is mostly because current photonic platforms lack some of the building blocks that are common in electronics, such as logic gates, high-level compilers and assemblers, analog-digital-analog conversion, and importantly, memory.

In some ML and neuromorphic applications (e.g. deep learning inference) the synaptic weights, once trained, do not have to be updated often or at all. In these cases, non-volatile analogue memory would be beneficial, a potential candidate being "in-memory" computing which can be implemented optically or electronically with PCMs [34, 66]. When controlled by digital electronic drivers running photonics-compatible firmware, one can implement a neural network running in real-time, precompute the necessary weights, and directly load an inference task onto hardware. Although, weights may not need to be updated often, there are cases (e.g. in long short-term memory recurrent neural networks) in which the output of the neurons need to be stored temporarily i.e. written to and read from memory fast. In these cases, PCMs are probably not the right technology to store that type of data. Digital—or perhaps short-term analog [111, 112]—electronic memory with electro-optic interfaces to analog photonics would be best suited to store that particular type of data without displacing a role for PCMs in storing long-term weights. When compared to digital memory, analog memory would be limited in precision and noise, but it has been shown [113] that deep and recurrent neural networks work well even in low precision.

All modern computers have heterogeneous memory technologies (registers, caches, flash, DRAM, etc.) within a single system. Neuromorphic photonic systems are expected to follow a similar principle with tighter co-packaging of electronic memory. We could also witness integration of novel photonic memory technologies. For example, moving beyond inference machines could benefit from non-volatile, yet reconfigurable, optical materials for memory. The concept of memory here is closely related to *learning*. In applications such as online learning and training, synaptic weights need to be updated frequently. In these cases, fast memory, normally provided by e.g. DRAM is necessary and where photonics DACs could play a role. In modern computers, memory is used to store both programs and data. The CPU juggles between reading instructions, executing them, and manipulating data in and out of a memory unit. Non-volatile optical materials could serve a role beyond just data storage or instruction storage. They can be part of the computational algorithm running on the neuromorphic processor. It is often thought that synaptic weights are effectively the long-term memory of a network, but physiological synapses are constantly changing according to neural activity and due to their molecular chemistry.

Non-volatile photonic memory can be implemented by cladding waveguides with PCMs, including chalcogenides, currently used for memory storage applications (such as rewritable optical disks DVD-RW) [35, 114]. These materials feature multiple stable phases of matter that have distinct optical properties. The most commonly used PCM currently is the archetypal alloy $Ge_2Sb_2Te_5$, but other compositions are also being investigated (e.g. the "low-loss" material GeSbSeTe [115]). Changing the state of the PCM (from amorphous to crystalline) varies the waveguide's effective refractive index (changing the optical-path length) and modulates the absorption. Since devices like silicon weight banks rely on the refractive index of a waveguide – as reviewed in Fig. 1 and rendered in Fig. 5 – PCM-cladded waveguides offer non-volatile weights that can be reconfigured with optical or electrical signals [34, 116]. The process of setting the weights is also reversible, which limits the need to read from and write to electronic memories with DACs and ADCs. Non-volatile weights for photonic neural networks can also be achieved in a cryogenic setting through the long-term storage of superconducting currents [43]. With a reversible weight-setting process, dynamic synaptic plasticity and online learning can be enabled by local feedback circuits implemented in each neuron.

**Correcting for variability.** Analog circuits often need trimming to function correctly and correct for manufacturing variabilities and environmental sensitivity. In integrated photonics, resonant devices, such as microring resonators, can pose particular issues in this respect [117]. One approach to address this challenge is resonance trimming. Active trimming, for example inducing changes in refractive index by heating a waveguide, is useful to counteract environmental variability, such as temperature and vibration, but requires constant input power. Furthermore, it requires microsecond response speeds and a continuous, rather than discrete, response to an electrical signal. An alternative approach is to use permanent or non-volatile methods to trim the refractive index of a device, i.e. passive trimming. This approach can be employed to correct for fabrication variation, or to "preprogram" a circuit to a default state. Permanent methods, typically employed during manufacture, include electron beam induced compaction and strain of oxide cladding [118], electron beam bleaching of a polymer cladding [119] and, the most CMOS foundry compatible approach, germanium ion implantation and annealing [120]. Non-volatile methods include field-programmable PCMs, and can be reconfigured in-place [35].

In order to map application tasks to photonic hardware, new analog-aware compilers are necessary. Compilers on conventional computers change high-level code to machine instructions that differ depending on the computer. Likewise, photonic compilers should be able to abstract away the idiosyncrasies that happen when we represent signals in WDM lightwaves subject to nonlinear distortion, limited dynamic range, limited gain, and crosstalk. This work is in progress in the academic community in the fields of neuromorphic photonics [7, 12, 14, 30] and programmable photonics [121, 122].

**Frequency comb-based WDM sources.** In combination with on-chip multiplexers and demultiplexers for WDM, tailored light sources which provide evenly spaced emission wavelengths aligned to standardized communication channels are desirable. While one approach is to use an array of integrated WDM lasers, an attractive alternative is chipscale frequency combs leveraging nonlinear optics, requiring only one laser.

Using soliton microcombs, frequency combs can now be built using CMOS-compatible photonic integrated circuits [123]. The resonators are designed such that the nonlinearity compensates for the dispersion and a single soliton can circulate within the cavity. A frequency comb then forms inside the resonator with a frequency spacing 13 equal to the free spectral range (FSR) of the resonator [124, 125]. A single soliton state is prepared in a high-Q ring resonator, leading to a very stable and broadband frequency comb with a wavelength spacing that can be aligned to the telecom ITU grid (100 GHz), determined by the FSR of the resonator. For photonic neuromorphic processing, chipscale microcombs provide the means to generate tailored input signals in a highly parallel fashion from a single source.

**Lithium niobate on insulator modulators.** Lithium niobate ($LiNbO_3$) was one of the early enabling platforms for realizing "planar lightwave circuits." $LiNbO_3$ offers attractive material properties, including a wide transparency window covering visible to mid infrared wavelengths and a strong electro-optical coefficient [126]. Lithium niobate also provides high second order optical nonlinearity which allows for parametric wavelength conversion and nonlinear mixing. Thin-film lithium niobate on insulator (LNOI) has emerged as a photonic substrate and become commercially available. The LNOI platform allows for building photonic circuits with dimensions compatible with silicon photonic devices [127] and provides the ability to integrate fast electro optical modulators and efficient nonlinear optical elements on the same chip.

LNOI-based modulators can achieve very high modulation frequencies (up to 110 GHz [128]) with low voltage-length product ($V_\pi L$), which is beneficial in terms of energy consumption and device footprint. Recent demonstrations showed monolithically integrated lithium niobate electro-optic modulators that feature CMOS-compatible driving voltage, data rates up to 210 Gb/s and an on-chip optical loss of less than 0.5 dB [129]. In combination with the capability to electro-optically modulate on chip, this approach enables the integration of nanophotonic waveguides, microring resonators, filters and modulators on the same chip. Such devices could lead to large-scale ultra-low-loss photonic circuits that are reconfigurable on a picosecond timescale.

**Photonic DACs.** Interfacing analog processors invariably requires ADC and DAC circuits operating at full data rate (tens of gigahertz). Going from digital-electronic to analog-photonic signals requires two rather costly conversion steps: D/A conversion and electro-optic modulation. One possibility is to combine these two steps into one, using photonic DACs to achieve high sampling rates, high precision, and low distortion, while being less affected by jitter or electromagnetic noise than electronic counterparts. Neuromorphic photonic processors would benefit from photonic DACs compatible with silicon photonic integration for reduced footprint, high sampling rates and low power consumption.

One approach is based on optical intensity weighting of multiwavelength signal, modulated with silicon MRRs with depletion-mode PN junctions [130, 131]. In this approach, the bit number is determined by the number of cascaded MRRs, which is further determined by the FSR of the MRR and the channel spacing each MRR occupies (which affects the crosstalk). The main challenge here is achieving high-speed operation with a high extinction ratio. A 2-bit photonic DAC was demonstrated in [130], with a silicon MRR with modulation data rate up to 128 Gb/s (64 Gbaud). Another 2-bit DAC demonstration was based on an SOI traveling-wave multi-electrode Mach-Zehnder modulator operating at 100 Gb/s (50 Gbaud) [131]. Scaling to a higher number of bits may be achieved with a coherent parallel photonic DAC proposed in [132].

## Conclusion

Neuromorphic photonics is the creation of optoelectronic hardware that is isomorphic to neural networks. As a consequence of this isomorphism, photonic neural networks will have remarkable capabilities, they will have strong technological and societal demand, and they can leverage existing algorithmic methodologies for programming and training, including the leaps and bounds occurring in the practice of "deep learning."

Research in photonic neural networks has multiplied considerably in recent years. Already under investigation are several architectural concepts with a variety of—not just implementations but—neuron models, training techniques, and topologies. This diversity implies that neuromorphic photonics research is not be expected to converge to a single winning implementation or a single application. Continuous research is required to identify applications where photonics will most excel over the continually advancing state-of-the-art in electronic computing. Relatedly, there will be a persistent demand for benchmarks comparing emerging photonic and electronic technologies. Most promising will be real-time applications where decisions must occur in a very short time. Moving forward, it is now time to focus on scaling the number of neurons integrated in single networks. With key photonic libraries having been demonstrated on scalable (silicon) photonics platforms, the critical technological challenges are co-packing of control electronics and light sources. Abetted by modern integrated platforms for programmable photonics, new ideas and devices for on-chip cascadability and nonlinearity, seemingly insurmountable barriers to the bandwidth of neuromorphic electronics, and an obvious societal demand for neural network processors, neuromorphic photonics has a high potential to extend the frontiers of machine learning and information processing.

## References

[1] LeCun, Y., Bengio, Y. & Hinton, G. Deep learning. *Nature* 521, 436–444 (2015).
[2] Wu, Y. *et al.* Google's neural machine translation system: Bridging the gap between human and machine translation (2016). arXiv:1609. 08144.
[3] Capper, D. *et al.* Dna methylation-based classification of central nervous system tumours. *Nature* 555, 469–474 (2018).
[4] Merolla, P. A. *et al.* A million spiking-neuron integrated circuit with a scalable communication network and interface. *Science* 345, 668–673 (2014).
[5] Davies, M. *et al.* Loihi: A neuromorphic manycore processor with on-chip learning. *IEEE Micro* 38, 82–99 (2018).
[6] Keyes, R. W. Optical logic-in the light of computer technology. *Optica Acta: International Journal of Optics* 32, 525–535 (1985).
[7] Prucnal, P. R. & Shastri, B. J. *Neuromorphic Photonics* (CRC Press, Boca Raton, FL, 2017).
[8] Magesan, E., Gambetta, J. M., Corcoles, A. D. & Chow, J. M. Machine learning for discriminating quantum measurement trajectories and ´improving readout. *Phys. Rev. Lett.* 114, 200501 (2015).
[9] Radovic, A. *et al.* Machine learning at the energy and intensity frontiers of particle physics. *Nature* 560, 41–48 (2018).

[10] Duarte, J. *et al.* Fast inference of deep neural networks in FPGAs for particle physics. *Journal of Instrumentation* 13, P07027–P07027 (2018).
[11] Kates-Harbeck, J., Svyatkovskiy, A. & Tang, W. Predicting disruptive instabilities in controlled fusion plasmas through deep learning. *Nature* 568, 526–531 (2019).
[12] Ferreira de Lima, T. *et al.* Machine learning with neuromorphic photonics. *Journal of Lightwave Technology* 37, 1515–1534 (2019).
[13] Han, J., Jentzen, A. & E, W. Solving high-dimensional partial differential equations using deep learning. *Proceedings of the National Academy of Sciences* 115, 8505–8510 (2018).
[14] Shen, Y. *et al.* Deep learning with coherent nanophotonic circuits. *Nat. Photon.* 11, 441–446 (2017).
[15] Jouppi, N. P. *et al.* In-datacenter performance analysis of a tensor processing unit. In *Proceedings of the 44th Annual International Symposium on Computer Architecture*, ISCA '17, 1–12 (Association for Computing Machinery, New York, NY, USA, 2017).
[16] Hughes, T. W., Minkov, M., Shi, Y. & Fan, S. Training of photonic neural networks through in situ backpropagation and gradient measurement. *Optica* 5, 864–871 (2018).
[17] Tait, A. N. *et al.* Demonstration of multivariate photonics: Blind dimensionality reduction with integrated photonics. *Journal of Lightwave Technology* 37, 5996–6006 (2019).
[18] Huang, C. *et al.* Demonstration of photonic neural network for fiber nonlinearity compensation in long-haul transmission systems. In *Optical Fiber Communication Conference*, Th4C–6 (Optical Society of America, 2020).
[19] Zhang, S. *et al.* Field and lab experimental demonstration of nonlinear impairment compensation using neural networks. *Nature communications* 10, 1–8 (2019).
[20] Kravtsov, K. S., Fok, M. P., Prucnal, P. R. & Rosenbluth, D. Ultrafast all-optical implementation of a leaky integrate-and-fire neuron. *Opt. Express* 19, 2133–2147 (2011).
[21] Tait, A. N., Nahmias, M. A., Shastri, B. J. & Prucnal, P. R. Broadcast and weight: An integrated network for scalable photonic spike processing. *J. Lightwave Technol.* 32, 4029–4041 (2014).
[22] Shainline, J. M., Buckley, S. M., Mirin, R. P. & Nam, S. W. Superconducting optoelectronic circuits for neuromorphic computing. *Phys. Rev. Applied* 7, 034013 (2017).
[23] Bangari, V. *et al.* Digital electronics and analog photonics for convolutional neural networks (deap-cnns). *IEEE Journal of Selected Topics in Quantum Electronics* 26, 1–13 (2020).
[24] Feldmann, J., Youngblood, N., Wright, C. D., Bhaskaran, H. & Pernice, W. H. P. All-optical spiking neurosynaptic networks with self-learning capabilities. *Nature* 569, 208–214 (2019).
[25] Goodman, J. W., Dias, A. R. & Woody, L. M. Fully parallel, high-speed incoherent optical method for performing discrete fourier transforms. *Opt. Lett.* 2, 1–3 (1978).
[26] Goodman, J. W., Leonberger, F. J., Kung, S.-Y. & Athale, R. A. Optical interconnections for vlsi systems. *Proceedings of the IEEE* 72, 850–866 (1984).
[27] Miller, D. A. B. Rationale and challenges for optical interconnects to electronic chips. *Proceedings of the IEEE* 88, 728–749 (2000).
[28] Psaltis, D. & Farhat, N. Optical information processing based on an associative-memory model of neural nets with thresholding and feedback. *Opt. Lett.* 10, 98–100 (1985).
[29] Soref, R. & Bennett, B. Electrooptical effects in silicon. *IEEE Journal of Quantum Electronics* 23, 123–129 (1987).
[30] Tait, A. N. *et al.* Neuromorphic photonic networks using silicon photonic weight banks. *Sci. Rep.* 7, 7430 (2017).
[31] Bogaerts, W. & Chrostowski, L. Silicon photonics circuit design: Methods, tools and challenges. *Laser & Photonics Reviews* 12, 1700237 (2018).
[32] Nozaki, K. *et al.* Femtofarad optoelectronic integration demonstrating energy-saving signal conversion and nonlinear functions. *Nature Photonics* 13, 454–459 (2019).
[33] Nahmias, M. A. *et al.* Photonic Multiply-Accumulate Operations for Neural Networks. *IEEE Journal of Selected Topics in Quantum Electronics* 26, 1–18 (2020).
[34] Ríos, C. *et al.* In-memory computing on a photonic platform. *Science advances* 5, eaau5759 (2019).
[35] Ríos, C. *et al.* Integrated all-photonic non-volatile multi-level memory. *Nature Photonics* 9, 725–732 (2015).
[36] Lecun, Y., Bottou, L., Bengio, Y. & Haffner, P. Gradient-based learning applied to document recognition. *Proceedings of the IEEE* 86, 2278–2324 (1998).
[37] Tait, A., Ferreira de Lima, T., Nahmias, M., Shastri, B. & Prucnal, P. Continuous calibration of microring weights for analog optical networks. *Photonics Technol. Lett.* 28, 887–890 (2016).
[38] Tait, A. N. *et al.* Microring weight banks. *IEEE J. Sel. Top. Quantum Electron.* 22 (2016).
[39] Shi, B., Calabretta, N. & Stabile, R. Deep neural network through an inp soa-based photonic integrated cross-connect. *IEEE Journal of Selected Topics in Quantum Electronics* 26, 1–11 (2020).

[40] Xu, X. *et al.* Photonic perceptron based on a kerr microcomb for high-speed, scalable, optical neural networks. *Laser* & *Photonics Reviews* n/a, 2000070.
[41] Reck, M., Zeilinger, A., Bernstein, H. J. & Bertani, P. Experimental realization of any discrete unitary operator. *Phys. Rev. Lett.* 73, 58–61 (1994).
[42] Carolan, J. *et al.* Universal linear optics. *Science* 349, 711 (2015).
[43] Shainline, J. M. *et al.* Superconducting optoelectronic loop neurons. *Journal of Applied Physics* 126, 044902 (2019).
[44] Chiles, J., Buckley, S. M., Nam, S. W., Mirin, R. P. & Shainline, J. M. Design, fabrication, and metrology of 10x100 multi-planar integrated photonic routing manifolds for neural networks. *APL Photonics* 3, 106101 (2018).
[45] Buckley, S. *et al.* All-silicon light-emitting diodes waveguide-integrated with superconducting single-photon detectors. *Applied Physics Letters* 111, 141101 (2017).
[46] Harris, N. C. *et al.* Efficient, compact and low loss thermo-optic phase shifter in silicon. *Opt. Express* 22, 10487–10493 (2014).
[47] Jayatilleka, H. *et al.* Wavelength tuning and stabilization of microring-based filters using silicon in-resonator photoconductive heaters. *Optics Express* 23, 25084 (2015).
[48] Tait, A. N. *et al.* Feedback control for microring weight banks. *Opt. Express* 26, 26422–26443 (2018).
[49] Patel, D. *et al.* Design, analysis, and transmission system performance of a 41 GHz silicon photonic modulator. *Optics Express* 23, 14263 (2015).
[50] Komljenovic, T. *et al.* Heterogeneous silicon photonic integrated circuits. *J. Lightwave Technol.* 34, 20–35 (2016).
[51] He, M. *et al.* High-performance hybrid silicon and lithium niobate Mach–Zehnder modulators for 100 Gbit s $^{-1}$ and beyond. *Nature Photonics* 13, 359–364 (2019).
[52] Sorianello, V. *et al.* Graphene-silicon phase modulators with gigahertz bandwidth. *Nature Photonics* 12, 40–44 (2018).
[53] Gholipour, B. *et al.* Amorphous metal-sulphide microfibers enable photonic synapses for brain-like computing. *Advanced Optical Materials* 3, 635–641 (2015).
[54] Goodman, J. W. Fan-in and fan-out with optical interconnections. *Optica Acta: International Journal of Optics* 32, 1489–1496 (1985).
[55] Nahmias, M. A., Shastri, B. J., Tait, A. N. & Prucnal, P. R. A leaky integrate-and-fire laser neuron for ultrafast cognitive computing. *IEEE J. Sel. Top. Quantum Electron.* 19, 1–12 (2013).
[56] Romeira, B. *et al.* Excitability and optical pulse generation in semiconductor lasers driven by resonant tunneling diode photo-detectors. *Opt. Express* 21, 20931–20940 (2013).
[57] Tait, A. N. *et al.* Silicon photonic modulator neuron. *Physical Review Applied* 11, 064043– (2019).
[58] Amin, R. *et al.* ITO-based electro-absorption modulator for photonic neural activation function. *APL Materials* 7, 081112 (2019).
[59] George, J. K. *et al.* Neuromorphic photonics with electro-absorption modulators. *Opt. Express* 27, 5181–5191 (2019).
[60] Nahmias, M. A. *et al.* An integrated analog O/E/O link for multi-channel laser neurons. *Applied Physics Letters* 108 (2016).
[61] Williamson, I. A. D. *et al.* Reprogrammable electro-optic nonlinear activation functions for optical neural networks. *IEEE Journal of Selected Topics in Quantum Electronics* 26, 1–12 (2020).
[62] McCaughan, A. N. *et al.* A superconducting thermal switch with ultrahigh impedance for interfacing superconductors to semiconductors. *Nature Electronics* 2, 451–456 (2019).
[63] Mourgias-Alexandris, G. *et al.* An all-optical neuron with sigmoid activation function. *Opt. Express* 27, 9620–9630 (2019). 17
[64] Hill, M., Frietman, E. E. E., de Waardt, H., Khoe, G.-D. & Dorren, H. All fiber-optic neural network using coupled SOA based ring lasers. *IEEE Trans. Neural Networks* 13, 1504–1513 (2002).
[65] Rosenbluth, D., Kravtsov, K., Fok, M. P. & Prucnal, P. R. A high performance photonic pulse processing device. *Optics Express* 17, 22767–22772 (2009).
[66] Sebastian, A. *et al.* Tutorial: Brain-inspired computing using phase-change memory devices. *Journal of Applied Physics* 124, 111101 (2018).
[67] Selmi, F. *et al.* Relative refractory period in an excitable semiconductor laser. *Phys. Rev. Lett.* 112, 183902 (2014).
[68] Peng, H. T. *et al.* Neuromorphic photonic integrated circuits. *IEEE Journal of Selected Topics in Quantum Electronics* 24, 1–15 (2018).
[69] Romeira, B., Avo, R., Figueiredo, J. M. L., Barland, S. & Javaloyes, J. Regenerative memory in time-delayed neuromorphic photonic ´ resonators. *Scientific Reports* 6, 19510 (2016).
[70] Shastri, B. J. *et al.* Spike processing with a graphene excitable laser. *Scientific Reports* 6, 19126 (2016).

[71] Coomans, W., Gelens, L., Beri, S., Danckaert, J. & Van der Sande, G. Solitary and coupled semiconductor ring lasers as optical spiking neurons. *Phys. Rev. E* 84, 036209 (2011).
[72] Brunstein, M. *et al.* Excitability and self-pulsing in a photonic crystal nanocavity. *Physical Review A* 85, 031803 (2012).
[73] Robertson, J., Deng, T., Javaloyes, J. & Hurtado, A. Controlled inhibition of spiking dynamics in vcsels for neuromorphic photonics: theory and experiments. *Opt. Lett.* 42, 1560–1563 (2017).
[74] Hopfield, J. J. Neural networks and physical systems with emergent collective computational abilities. *Proceedings of the National Academy of Sciences* 79, 2554–2558 (1982).
[75] Stewart, T. C. & Eliasmith, C. Large-scale synthesis of functional spiking neural circuits. *Proceedings of the IEEE* 102, 881–898 (2014).
[76] Bueno, J. *et al.* Reinforcement learning in a large-scale photonic recurrent neural network. *Optica* 5, 756–760 (2018).
[77] Lin, X. *et al.* All-optical machine learning using diffractive deep neural networks. *Science* (2018).
[78] Zuo, Y. *et al.* All-optical neural network with nonlinear activation functions. *Optica* 6, 1132–1137 (2019).
[79] Xu, S., Wang, J., Wang, R., Chen, J. & Zou, W. High-accuracy optical convolution unit architecture for convolutional neural networks by cascaded acousto-optical modulator arrays. *Opt. Express* 27, 19778–19787 (2019).
[80] Mehrabian, A., Miscuglio, M., Alkabani, Y., Sorger, V. J. & El-Ghazawi, T. A winograd-based integrated photonics accelerator for convolutional neural networks. *IEEE Journal of Selected Topics in Quantum Electronics* 26, 1–12 (2020).
[81] McMahon, P. L. *et al.* A fully programmable 100-spin coherent ising machine with all-to-all connections. *Science* 354, 614–617 (2016).
[82] Roques-Carmes, C. *et al.* Heuristic recurrent algorithms for photonic ising machines. *Nature Communications* 11, 249 (2020).
[83] Tang, P. T. P., Lin, T.-H. & Davies, M. Sparse Coding by Spiking Neural Networks: Convergence Theory and Computational Results. *arXiv:1705.05475* (2017).
[84] Davies, M. Benchmarks for progress in neuromorphic computing. *Nature Machine Intelligence* 1, 386–388 (2019).
[85] Maass, W. Networks of spiking neurons: The third generation of neural network models. *Neural Networks* 10, 1659 – 1671 (1997).
[86] Peng, H. *et al.* Temporal information processing with an integrated laser neuron. *IEEE Journal of Selected Topics in Quantum Electronics* 26, 1–9 (2020).
[87] Robertson, J., Hejda, M., Bueno, J. & Hurtado, A. Ultrafast optical integration and pattern classification for neuromorphic photonics based on spiking VCSEL neurons. *Scientific Reports* 10, 6098 (2020).
[88] Chakraborty, I., Saha, G., Sengupta, A. & Roy, K. Toward fast neural computing using all-photonic phase change spiking neurons. *Scientific Reports* 8, 12980 (2018).
[89] Fok, M. P., Tian, Y., Rosenbluth, D. & Prucnal, P. R. Pulse lead/lag timing detection for adaptive feedback and control based on optical spike-timing-dependent plasticity. *Opt. Lett.* 38, 419–421 (2013).
[90] Toole, R. *et al.* Photonic implementation of spike-timing-dependent plasticity and learning algorithms of biological neural systems. *Journal of Lightwave Technology* 34, 470–476 (2016).
[91] Xiang, S. *et al.* STDP-based unsupervised spike pattern learning in a photonic spiking neural network with VCSELs and VCSOAs. *IEEE Journal of Selected Topics in Quantum Electronics* 25, 1–9 (2019).
[92] Larger, L. *et al.* Photonic information processing beyond turing: an optoelectronic implementation of reservoir computing. *Opt. Express* 20, 3241–3249 (2012).
[93] Paquot, Y. *et al.* Optoelectronic reservoir computing. *Scientific Reports* 2, 287 (2012).
[94] Duport, F., Schneider, B., Smerieri, A., Haelterman, M. & Massar, S. All-optical reservoir computing. *Opt. Express* 20, 22783–22795 (2012).
[95] Brunner, D., Soriano, M. C., Mirasso, C. R. & Fischer, I. Parallel photonic information processing at gigabyte per second data rates using transient states. *Nat Commun* 4, 1364 (2013).
[96] Vandoorne, K. *et al.* Experimental demonstration of reservoir computing on a silicon photonics chip. *Nat. Commun.* 5 (2014).
[97] Brunner, D. *et al. Tutorial: Photonic neural networks in delay systems*, vol. 124 (2018).
[98] Brunner, D., Soriano, M. C. & der Sande, G. V. *Photonic Reservoir Computing* (De Gruyter, Berlin, Boston, 08 Jul. 2019).
[99] Antonik, P., Marsal, N., Brunner, D. & Rontani, D. Human action recognition with a large-scale brain-inspired photonic computer. *Nature Machine Intelligence* 1, 530–537 (2019).
[100] Sun, C. *et al.* Single-chip microprocessor that communicates directly using light. *Nature* 528, 534–538 (2015).

[101] Stojanovic, V. *et al.* Monolithic silicon-photonic platforms in state-of-the-art CMOS SOI processes [Invited]. *Optics Express* 26, 13106 (2018).
[102] Jha, A. *et al.* Lateral bipolar junction transistor on a silicon photonics platform. *Opt. Express* 28, 11692–11704 (2020).
[103] Giewont, K. *et al.* 300-mm Monolithic Silicon Photonics Foundry Technology. *IEEE Journal of Selected Topics in Quantum Electronics* 25, 1–11 (2019).
[104] Zhou, Z., Yin, B. & Michel, J. On-chip light sources for silicon photonics. *Light: Science & Applications* 4, e358–e358 (2015).
[105] Song, B., Stagarescu, C., Ristic, S., Behfar, A. & Klamkin, J. 3d integrated hybrid silicon laser. *Opt. Express* 24, 10435–10444 (2020).
[106] Mack, M. *et al.* Luxtera's silicon photonics platform for transceiver manufacturing. In *2014 International Conference on Solid State Devices and Materials*, 506–507 (Luxtera, Inc., 2014).
[107] Billah, M. R. *et al.* Hybrid integration of silicon photonics circuits and inp lasers by photonic wire bonding. *Optica* 5, 876–883 (2018).
[108] Liang, D. & Bowers, J. E. Recent progress in lasers on silicon. *Nature Photonics* 4, 511–517 (2010).
[109] Chen, S. *et al.* Electrically pumped continuous-wave iii–v quantum dot lasers on silicon. *Nature Photonics* 10, 307–311 (2016).
[110] Xia, Q. *et al.* Roadmap on emerging hardware and technology for machine learning. *Nanotechnology* (2020).
[111] Ambrogio, S. *et al.* Equivalent-accuracy accelerated neural-network training using analogue memory. *Nature* 558, 60–67 (2018).
[112] Li, C. *et al.* Long short-term memory networks in memristor crossbar arrays. *Nature Machine Intelligence* 1, 49–57 (2019).
[113] Sze, V., Chen, Y., Yang, T. & Emer, J. S. Efficient processing of deep neural networks: A tutorial and survey. *Proceedings of the IEEE* 105, 2295–2329 (2017).
[114] Cheng, Z. *et al.* Device-level photonic memories and logic applications using phase-change materials. *Advanced Materials* 30, 1802435 (2018).
[115] Zhang, Y. *et al.* Broadband transparent optical phase change materials for high-performance nonvolatile photonics. *Nature Communications* 10, 4279 (2019).
[116] Cheng, Z., Ríos, C., Pernice, W. H. P., Wright, C. D. & Bhaskaran, H. On-chip photonic synapse. *Science Advances* 3 (2017).
[117] Bogaerts, W. *et al.* Silicon microring resonators. *Laser & Photonics Reviews* 6, 47–73 (2012).
[118] Schrauwen, J., Van Thourhout, D. & Baets, R. Trimming of silicon ring resonator by electron beam induced compaction and strain. *Optics Express* 16, 3738 (2008).
[119] Prorok, S., Petrov, A. Y., Eich, M., Luo, J. & Jen, A. K.-Y. Trimming of high-Q-factor silicon ring resonators by electron beam bleaching. *Optics Letters* 37, 3114 (2012).
[120] Milosevic, M. M. *et al.* Ion Implantation in Silicon for Trimming the Operating Wavelength of Ring Resonators. *IEEE Journal of Selected Topics in Quantum Electronics* 24, 1–7 (2018).
[121] Harris, N. C. *et al.* Linear programmable nanophotonic processors. *Optica* 5, 1623–1631 (2018).
[122] Perez, D., Gasulla, I., Mahapatra, P. D. & Capmany, J. Principles, fundamentals, and applications of programmable integrated photonics. *Adv. Opt. Photon.* 12, 709–786 (2020).
[123] Gaeta, A. L., Lipson, M. & Kippenberg, T. J. Photonic-chip-based frequency combs. *Nature Photonics* 13, 158–169 (2019).
[124] Kippenberg, T. J., Holzwarth, R. & Diddams, S. A. Microresonator-based optical frequency combs. *Science* 332, 555–559 (2011).
[125] Del'Haye, P. *et al.* Optical frequency comb generation from a monolithic microresonator. *Nature* 450, 1214–1217 (2007).
[126] Turner, E. H. High-frequency electro-optic coefficients of lithium niobate. *Applied Physics Letters* 8, 303–304 (1966).
[127] Wang, C., Zhang, M., Stern, B., Lipson, M. & Loncar, M. Nanophotonic lithium niobate electro-optic modulators. *Opt. Express* 26, 1547–1555 (2018).
[128] Mercante, A. J. *et al.* 110 ghz cmos compatible thin film linbo3 modulator on silicon. *Opt. Express* 24, 15590–15595 (2016).
[129] Wang, C. *et al.* Integrated lithium niobate electro-optic modulators operating at cmos-compatible voltages. *Nature* 562, 101–104 (2018).
[130] Sun, J. *et al.* A 128 gb/s pam4 silicon microring modulator with integrated thermo-optic resonance tuning. *Journal of Lightwave Technology* 37, 110–115 (2019).

[131] Patel, D., Samani, A., Veerasubramanian, V., Ghosh, S. & Plant, D. V. Silicon photonic segmented modulator-based electro-optic dac for 100 gb/s pam-4 generation. *IEEE Photonics Technology Letters* 27, 2433–2436 (2015).

[132] Meng, J., Miscuglio, M., George, J. K., Babakhani, A. & Sorger, V. J. Electronic Bottleneck Suppression in Next-generation Networks with Integrated Photonic Digital-to-analog Converters. *arXiv Optics* arXiv:1911.02511 (2019).

[133] Bogaerts, W. & Rahim, A. Programmable Photonics: An Opportunity for an Accessible Large-Volume PIC Ecosystem. *IEEE Journal of Selected Topics in Quantum Electronics* PP, 1–1 (2020).

**Acknowledgements**

B.J.S. acknowledges support from Natural Sciences and Engineering Research Council of Canada (NSERC). T.F.L. and P.R.P. acknowledge the support from Office of Naval Research (ONR), Defense Advanced Research Projects Agency (DARPA), and National Science Foundation (NSF). We thank J. Shainline, P. Kuo, and N. Sanford for editorial contributions.

**Competing interests**

The authors declare no competing interests.

**Additional information**

**Correspondence** should be addressed to B.J.S. or A.N.T.

**Box 1: Neuromorphic engineering.**

Neural network models used in engineering are much simpler than those describing biological neural networks, yet they offer a general computing framework for a wide range of problems. Each artificial neuron in a network can be considered as two functional blocks, pictured below: a weighted addition unit and a nonlinear unit. Weighted addition has multiple inputs that are "fanned-in", and one output representing a linear combination of the inputs. The nonlinear unit applies an "activation function" to the weighted sum (see Figure), yielding the output of the neuron. The output of the neuron is broadcast (or "fanned-out") to many other neurons, possibly including itself. In the Figure, we identify a few possible ways of building a network. The first is a feedforward neural network (as used in "deep" neural networks), meaning that signals travel from left to right. The second is a recurrent neural network, where each neuron can receive outputs from previous, subsequent, and the same layer—these are called *recurrent* connections. A *reservoir* can be constructed with neural networks containing random, but fixed recurrent connections. Another important class of neural networks that form an essential part of deep learning are *convolutional* neural networks which can be feedforward or recurrent.

While these models are constructed of simple elements, it was shown that networks of neurons are sufficient to perform sophisticated computations and tasks, as exemplified by the recent achievements of machine learning. Implementing neural network models directly in hardware, as research shows, leads to speeds and efficiencies unmatched by software implementations, while borrowing the immense body of knowledge about neural network modeling, design, and training. Engineering this architecture requires creating hardware that is isomorphic to neurons and neural networks—hence *neuromorphic*. When this isomorphism is achieved, the governing physical dynamics of the hardware will carry out the neural computations in an analog way. The figure below shows three isomorphisms that have been explored in photonics: a weighted network, a continuous-time neuron, and a spiking neuron. Each column matches up the model equations and diagram with some photonic devices whose behavior approximates the equations.

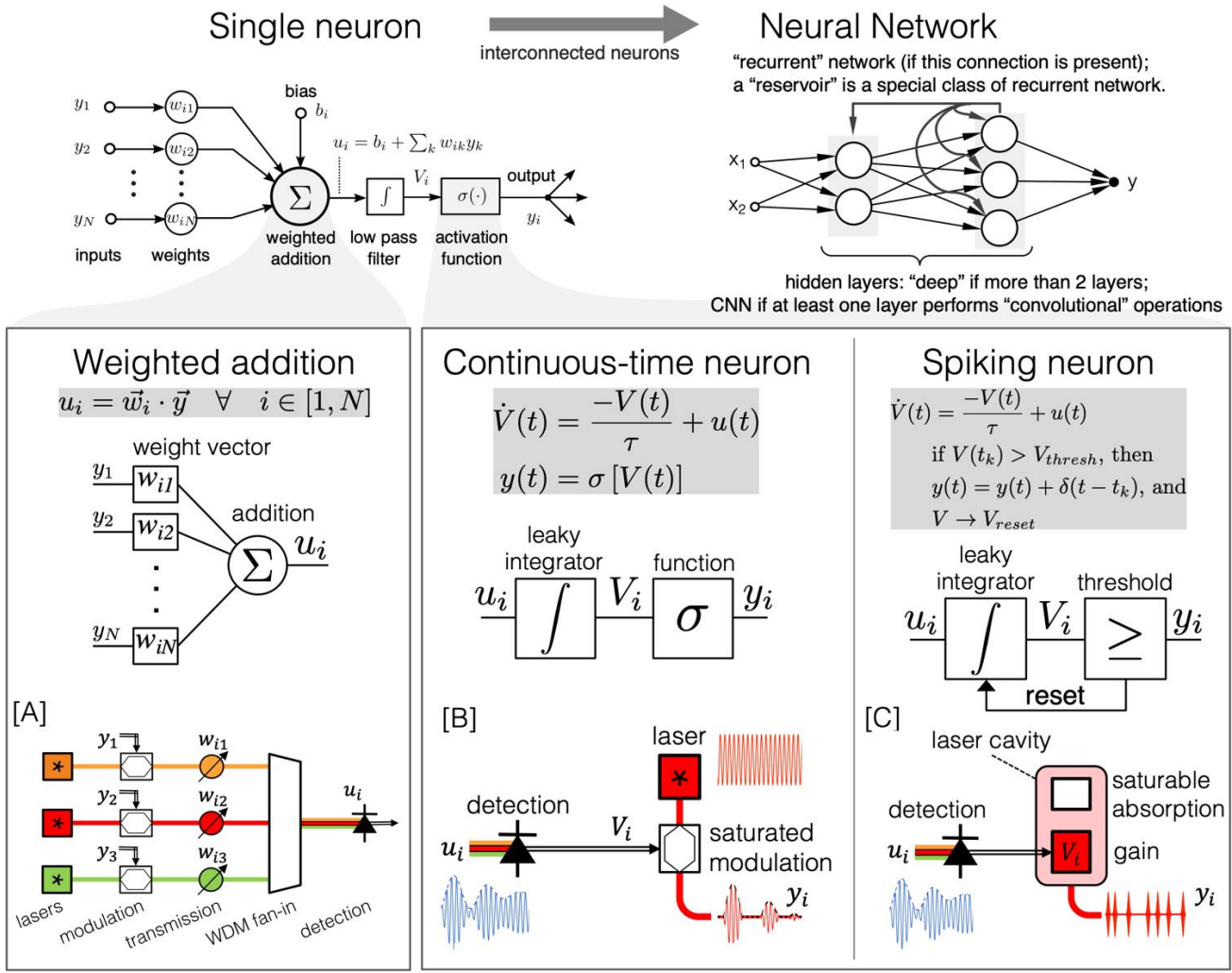

Citations for figure to be filled in once references are finalized. A=[14, 24, 30], B = [57, 61], C=[67, 70].

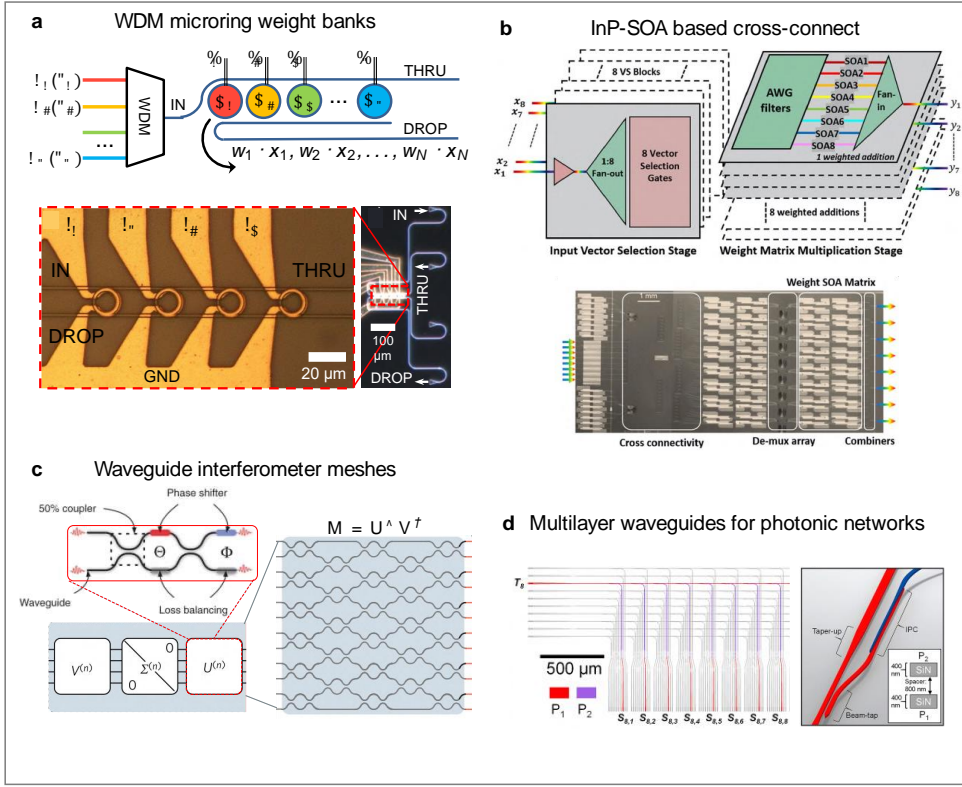
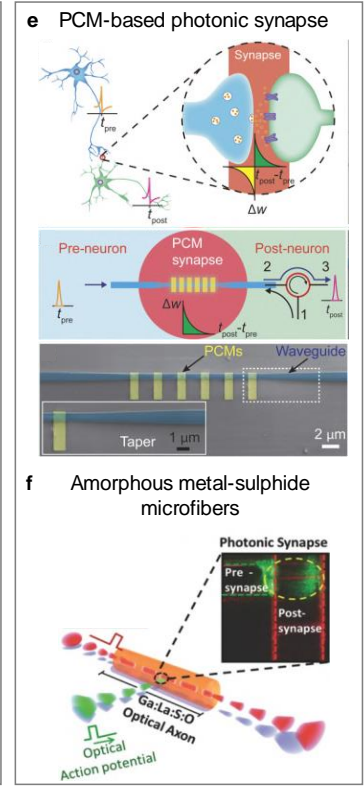

**Figure 1: Implementations of weights (photonic synapses).**

**a** Parallel weighting of WDM signals ($x_i$) with MRR weight banks as tunable filters [37, 38]. Optical micrograph shows fabricated silicon MRR weight bank with metal heaters for thermo-optic tuning of weights ($\omega_i$) with currents ($I_i$). Balanced photodetector sums these signals ($\Sigma\omega x$) and allows for positive and negative weights.

**b** Weighted addition with an SOA chip on an InP platform [39]. Schematic and microscope image of a chip co-integrating 8 weighted additions for 8 WDM input vectors and provides 8 WDM outputs.

**c** A MZI composed of waveguides and directional couplers with phase shifters implements a unitary transform [14]. Representing a weight matrix $M = U\Sigma V^\dagger$ through singular value decomposition, unitary matrices U and $V^\dagger$ are implemented with MZIs, and diagonal matrix $\Sigma$ with a Mach-Zehnder modulator.

**d** Photonic routing and weighting scheme for all-to-all connectivity using two vertically integrated planes of silicon nitride waveguides with a beam-tap and an inter-planar coupler (IPC). [44].

**e** Photonic synapse implemented with PCMs integrated on silicon nitride waveguides [116]. Synaptic weight is varied by the number of optical pulses sent down the waveguide.

**f** Photonic synapses demonstrated using metal-sulphide microfibers [53]. Transmission of pulses along the fiber is altered through photodarkening as a result of exposure at a sub-bandgap wavelength. This photomodulation plays the role of either inhibitory or excitatory action potentials in the postsynaptic axon.

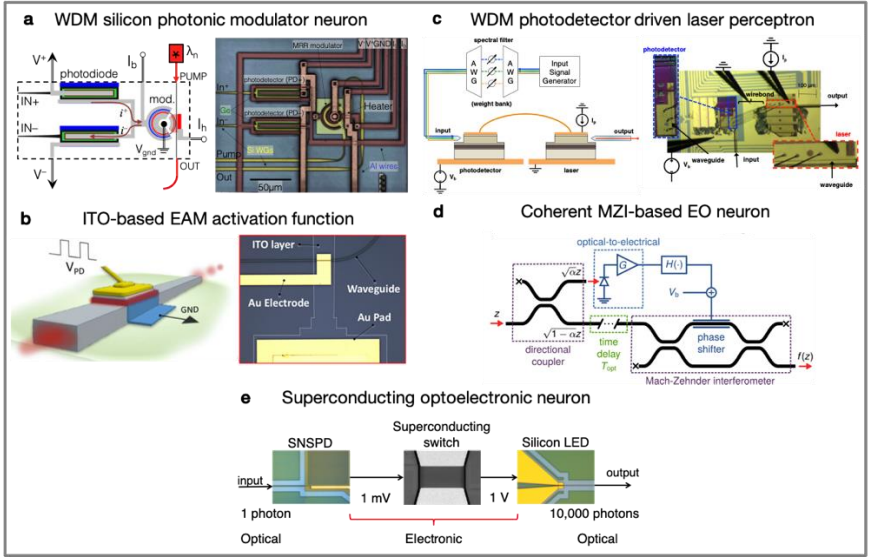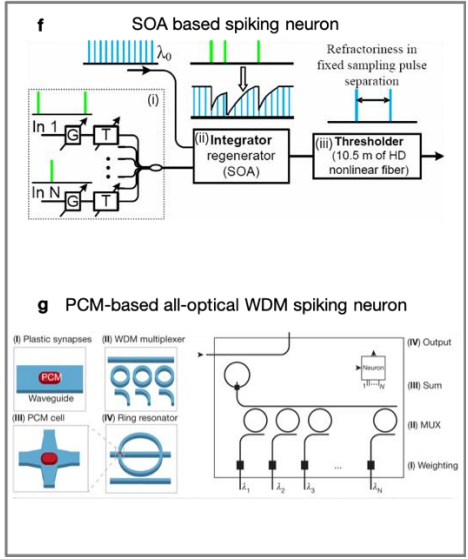

**Figure 2: Photonic neurons incorporating weighting and nonlinearity.**

**a** Silicon photonic modulator neuron [57]. A balanced photodiode (from Fig. 1a) sums multiple wavelengths and implements positive (excitatory) and negative (inhibitory) weights, and drives a ring modulator exploiting its electro-optic nonlinearity.

**b** Similar to (a) but an electro-absorption modulator (EAM) with an indium tin oxide (ITO) layer monolithically integrated into silicon photonic waveguides [58].

**c** The device utilizes WDM to achieve multi-channel fan-in, a photodetector to sum signals together, and a laser cavity to perform a nonlinear operation [60].

**d** A photodetector-driven MZI-based nonlinear activation function [61] for waveguide interferometer mesh weights (as in Fig. 1c).

**e** A superconducting optoelectronic spiking neuron based on a superconducting-nanowire single-photon detector (SNSPD) driving a superconducting switch (amplifier) [62] followed by a silicon LED [45].

**f** An integrate-and-fire semiconductor optical amplifier (SOA) spiking neuron [65]. Neuron inputs are weighted and delayed (with attenuators and delay lines), integrated with an SOA, and thresholded with a highly Ge-doped fiber.

**g** A PCM-based spiking neuron [24]. Inputs are weighted using PCM synapses (similar to Fig. 1e), summed using a WDM multiplexer (MUX), and thresholded with a PCM cell on a ring resonator.

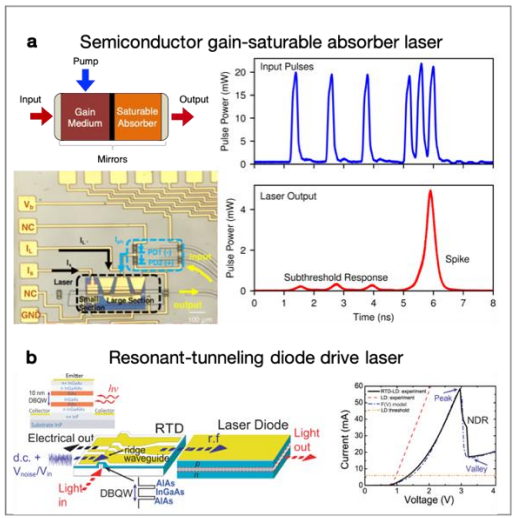
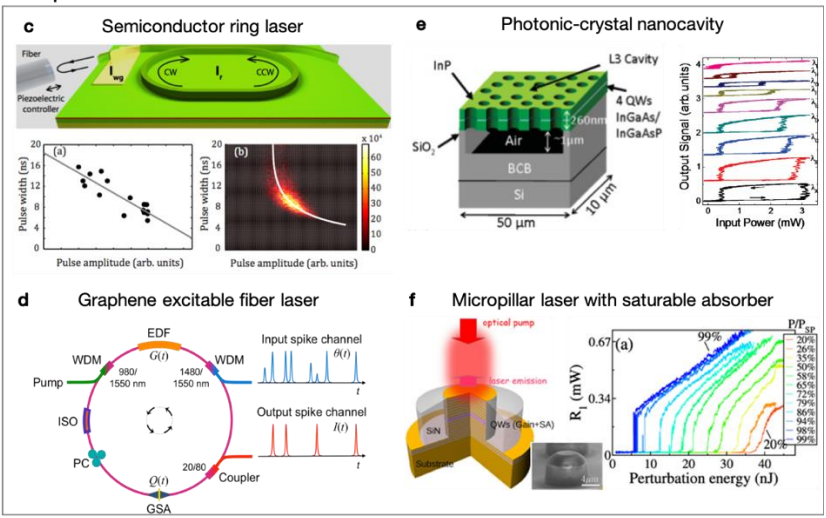

**Figure 3: Excitable lasers and resonators for spiking.**

**a** A two-section gain (integrate) and saturable absorber (SA; thresholder) excitable laser as an integrate-and-fire neuron [55] and micrograph of electrically injected excitable distributed feedback laser [68] driven by balanced photodetector pair. Plots show measured excitable dynamics of the laser.

**b** Resonant-tunneling diode (RTD) photodetector and laser diode [56]. Excitability is achieved by biasing a double barrier quantum well (DBQW) within the RTD in the negative differential resistance (NDR) region of its dc current–voltage curve.

**c** A semiconductor ring laser consisting of an electrically pumped III–V ring resonator coupled to a waveguide [71]. Two counterpropagating (CW and CCW) modes per frequency, lead to bistability. Excitability arises when this symmetry is broken.

**d** Graphene SA excitable fiber laser [70]. An erbium-doped fiber (EDF) acts as gain medium that is optically injected and pumped.

**e** InP-based two-dimensional photonic crystal nanocavity with quantum wells (QWs) [72]. This device exploits fast third-order nonlinearity for excitability. Hysteresis cycles show bistability with different detuning values with respect to the cavity resonance.

**f** Optically pumped III-V micropillar laser with a SA [67]. Plot shows amplitude response to a single pulse perturbation versus perturbation energy for bias pump $P$ relative to the self-pulsing threshold $P_{SP}$ demonstrating the distinction between an excitable and self-pulsing threshold.
19

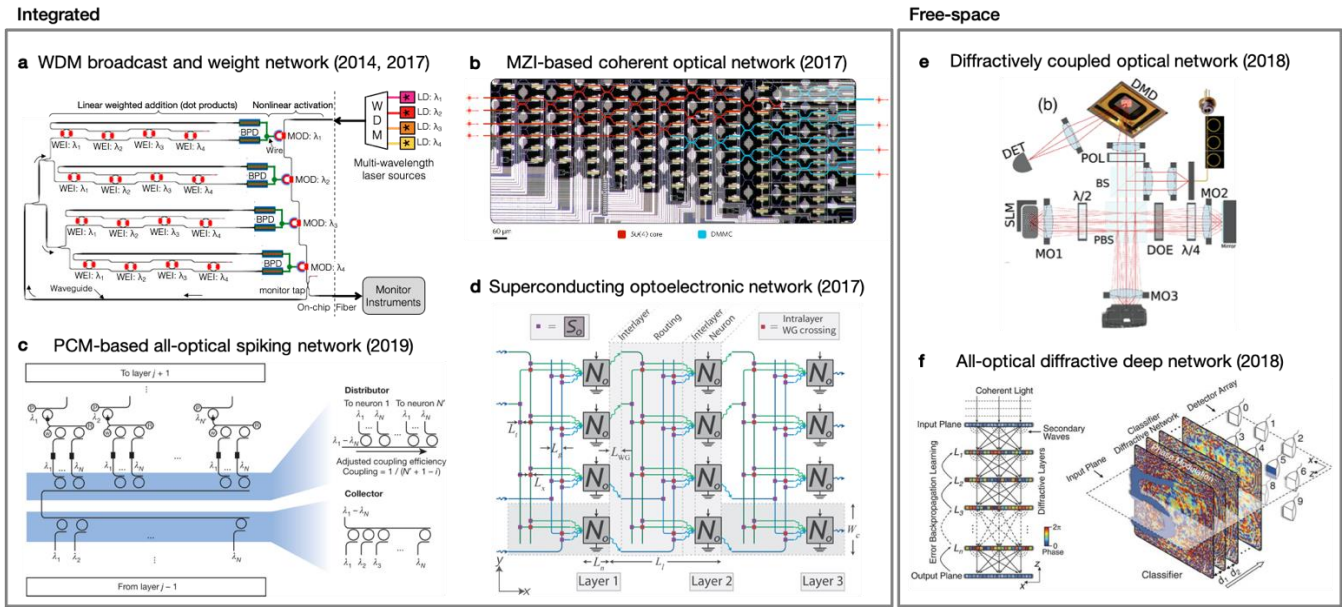

**Figure 4: Photonic neural network implementations.**

**a** A multiwavelength "broadcast-and-weight" photonic neural network [21, 57] composed of MRR weight banks (WEI) (as in Fig. 1a) and balanced photodiodes (BPD) for linear weighting and summing, and MRR modulators (MOD) for nonlinear activation (as in Fig. 2a).

**b** A coherent all-optical feed-forward network composed of MZI meshes (as in Fig. 1c) configured to implement a matrix using singular value decomposition: red meshes implement a unitary matrix and blue meshes implement a diagonal matrix [14].

**c** A WDM-based all-optical neural networks using PCM [24]. A collector made of MRRs multiplexes optical pulses from the previous layerand a distributor broadcasts the input signal equally to the PCM synapses of each neuron (in Fig. 2g).

**d** A multilayer perceptron implemented (as in Fig. 2e) with superconducting optoelectronic network platform [22] with $N_O$ neurons.

**e** Diffractively coupled photonic nodes forming a large-scale recurrent neural network [76]. A spatial light modulator (SLM) encodes the networks' state and a digital micromirror device (DMD) creates a spatially modulated image of SLM's state. The output is obtained via superimposing the detected modulated intensities.

**f** Diffractive deep neural network based on coherent waves [77]. It comprises multiple transmissive (or reflective) layers, where each point on a given layer acts as a neuron, with a complex-valued transmission (or reflection) coefficient.

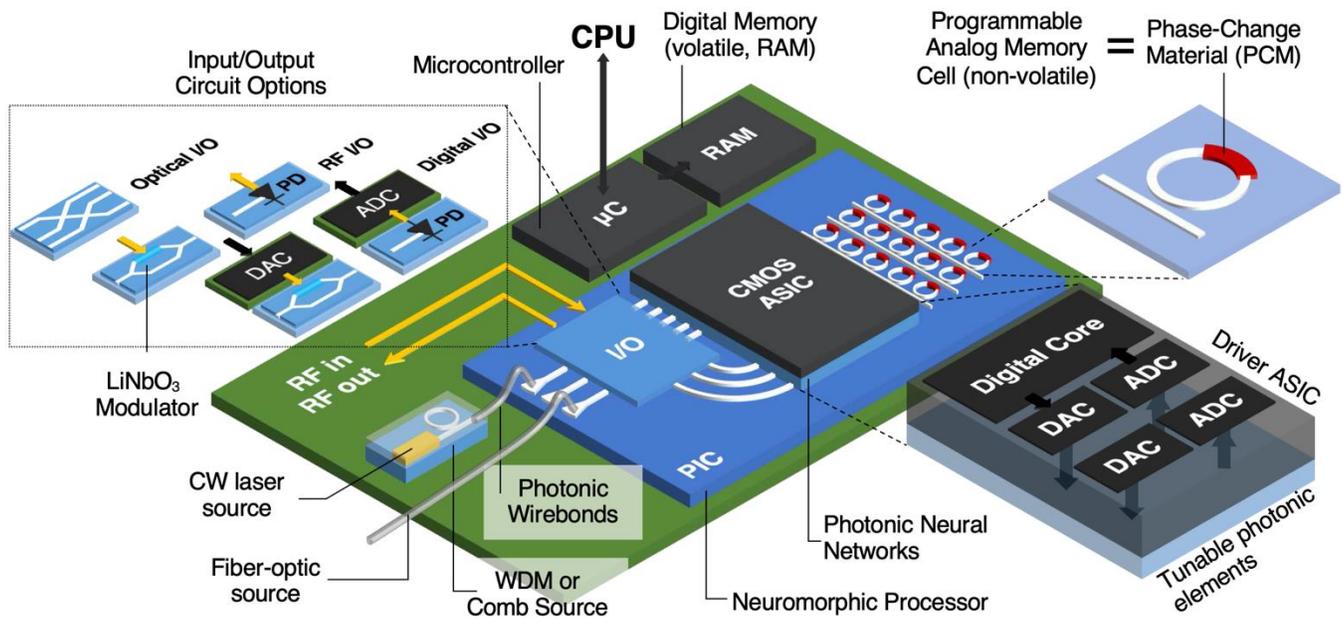

**Figure 5: Neuromorphic photonic processor architecture.** A concept system-in-package employing commercially available photonic packaging technology and some emerging ideas in the field of integrated photonics. At a high-level, the co-integration, packaging, and I/O strategies are closely related to those in programmable photonics [133]. The photonic neural network with configurable optical elements resides on a silicon photonic integrated circuit (PIC) die. Some elements can be configured in an analog, non-volatile way by phase change materials (PCM). White lines represent waveguide routing. One key challenge is getting optical power onto the silicon die. Optical power can be provided by an optically active die, i.e. able to generate light, or, alternatively, an external fiber interface. For signal I/O (top left inset), electrical-to-optical conversions and vice versa are performed by silicon photonic modulators (hexagonal structures) and photodetectors (triangles with crosses). This means that all package-level I/O can be electrical or optical, digital or analog, depending on the user's application. The other key challenge is controlling the photonic neural network. Black boxes represent CMOS dies satisfying various control, interface, and programming roles. The digital programming interface consists of a microcontroller (µC) with co-located digital memory (RAM), both of which are standard components. A CMOS application-specific integrated circuit (ASIC) is flip-chip bonded to the PIC. The ASIC generates a large number of voltages to drive the electrooptic elements (e.g. waveguide-embedded heaters) and thus configure a photonic neural network. The ASIC also provides digital-to-analog converters (DACs) to set drive voltages based on µC instructions and digital registers to maintain drive values when not being addressed by the µC. Since it is outside of the primary signal pathway, the ASIC need not be high-bandwidth and thus can be manufactured on modestly performant CMOS nodes using commercially available design blocks.